\documentclass[aps,prl,twocolumn,groupedaddress,showpacs]{revtex4}
\usepackage{epsf}
\def\etc{\hbox{\it etc.}}
\def\ie{\hbox{\it i.e.}}
\def\nn{\nonumber}
\def\beq{\begin{equation}}
\def\eeq{\end{equation}}
\def\bea{\begin{eqnarray}}
\def\eea{\end{eqnarray}}
\pacs{64.60.F-,05.70.Jk,64.60.ae,75.10.Hk}

\begin{document}
\title{Critical behavior of the Ising model with long range interactions}
\author{M. Picco}
\affiliation{LPTHE\footnote[1]{Unit\'e mixte de recherche du CNRS 
 UMR 7589.}, Universit\'e Pierre et Marie Curie - Paris 6,
                  4 place Jussieu, 75252 Paris cedex 05, France
	         }

\date{\today}

\begin{abstract}
\noindent

We present results of a Monte Carlo study for the  ferromagnetic Ising model with long range interactions 
in two dimensions. This model has been simulated for a large range of interaction parameter $\sigma$ and for large sizes. 
We observe that the results close to the change of regime from intermediate to short range do not agree with 
the renormalization group predictions. 
\end{abstract}

\maketitle


While the Ising model long range interactions (LRI)  has been studied for a long time as a generalization of the Ising model with short range interactions, 
very few results have been obtained for the case of weak interactions decaying faster than the dimension $d$ since the first works in the 70's \cite{mukamel}.
We will recall some of these results after defining the model that we will consider in this letter. 
The Ising model with LRI is defined by the Hamiltonian:
\beq
{\cal H} = -  \sum_{<ij>} {J\over r_{ij}^{d+\sigma}} S_i S_j   \; ,
\eeq
with spins taking values $S_i = \pm 1$ on the sites $i$ of a regular lattice and $r_{ij}$ the distance 
between the spins on the sites $i$ and $j$. The sum $<ij>$ is over all the pair of spins 
and we will consider only a ferromagnetic interaction $J>0$. 

In 1972, Fisher and al. \cite{fisher} performed a renormalization group (RG) study for the $O(n)$ model with LRI (the Ising model 
corresponding to $n=1$). 
They identified three regimes (for $d > 1$) :
i) the classical regime with $\sigma \leq d/2$ which is believed to be with a mean-field behavior, 
ii) an intermediate regime for $d/2 < \sigma < 2 $, 
iii) the short range regime for $\sigma \geq 2$.  
The value $\sigma=d/2$ marks the border of the mean field regime, meaning that the ordinary perturbation parameter 
$\epsilon = 4 -d $ is replaced by $2\sigma -d$. 
In \cite{fisher} the relation $\eta =2  - \sigma $ 
was also conjectured in the intermediate regime.
This result was questioned since for $\sigma = 2$, the exponent 
$\eta$ vanish while for $\sigma > 2$ its value has to be the one of the short range model $\eta_{sr}$ which is positive for $d < 4 $. 
Then it would imply a jump of the exponent $\eta$ from 
$0$ up to $\eta_{sr}$ at $\sigma=2$. This point was first considered by Sak \cite{sak}, 
who, by taking in account higher order terms in the RG calculations, predicted that 
the change of behavior from the intermediate to the short range regime takes place 
at $\sigma=  2- \eta_{sr}$.  
Many other studies have considered also this problem with various conclusions. In particular, 
van Enter \cite{van} obtained that for $n \ge 2 $, long range perturbations are relevant in the regime 
$2 - \eta_{sr} \leq \sigma \leq 2 $ in contradiction with Sak results. 
Gusm\~ao and Theumann \cite{gusthe}, by considering a development in terms of $\epsilon'=2\sigma - d$ in place 
of  $\epsilon = 2 - \sigma$, obtained a similar result, namely the stability of the long range perturbation for $\sigma \leq 2$. 

Note that all these studies are using a renormalisation group approach with an $\epsilon$ expansion of a Landau-Ginzburg effective Hamiltonian 
such that the propagator contains a $p^\sigma$ term in addition to the ordinary $p^2$ term. While it is know that this approach gives accurate results 
for two and three dimensions for the short range model (with just the $p^2$ term) \cite{jlgjzj}, it is only after comparing these predictions with other methods,
numerical, high temperature expansions, (and the exact result in two dimensions), \etc\ that we can believe in these predictions. Similar comparisons need 
also to be done when considering the case with LRI and this is the main purpose of the study presented in this letter. 

A first numerical study of the  exponent $\eta$ for $d=2$ as a  function of $\sigma$ has already been done by  Luijten and Bl\"ote \cite {lb}. 
In particular, they obtained in the  intermediate regime a result  well described by the 
exponent $\eta = 2- \sigma$ up to $2 - \sigma = \eta_{sr}$ and $ \eta = \eta_{sr} = 1/4$ for larger $\sigma$.
Thus their measured exponent is in agreement with $ \eta~=~\mbox{max}(2-\sigma, 1/4)$  which corresponds to the RG predictions of \cite{sak}. 


In the present study, we improve Luijten and Bl\"ote study. In particular, we  repeat the measurement of $\eta$ close 
to the region where its behavior is changing, {\it i.e.} for $\sigma \simeq 2 - \eta_{sr}$. We confirm their result that there is no discontinuity but 
we  measure a clear deviation from the behavior predicted by Sak \cite{sak}.


In order to be able to consider large lattices, we need to employ an efficient algorithm. 
Since the model is ferromagnetic, we can employ a cluster algorithm  which will reduce the auto correlation time, that is the number of 
upgrades between two successive independent configurations. For the short range Ising model, 
Wolff algorithm, which builds a single cluster per update, is the most efficient one \cite{wolff}. 
We will adapt this algorithm to the case with LRI. 
The algorithm can be summarized by the following steps. We start from a spin $S_i$ at some randomly selected position $i$. 
Next, we add to this spin any other spin $S_j$ with a probability 
\beq
\delta_{S_i,S_j} \left(1 - e^{-2 \beta {J\over r_{ij}^{d+\sigma}}}\right) = \delta_{S_i,S_j} (1 - p(r_{ij}))\; ,
\label{prob}
\eeq
with $\beta $ the inverse temperature.
We repeat the same operation with all the added spins. 
In order to build a cluster containing $M$ spins on a lattice of $N$ sites, we have to 
compute the probability (\ref{prob}) for $\simeq N \times M$ bonds, then the number of operations 
is $\mathcal{O}(N\times M)$. This number of operations will be drastically reduced by our algorithm that we 
will describe now. We will present here only the main steps of this algorithm, a more complete version will be presented elsewhere \cite{mp}. 

The main idea is that since $p(r_{ij})$ is very small for large $r_{ij}$, then it is much faster to compute the probability that one (or more than one spin) among all the spins at this distance are connected to the original spin $S_i$. In practice, 
to build this cluster we proceed as follow. 
First, starting from some arbitrary position $i$, we order in some pile ${\cal P}_1(k=1, \cdots, N-1)$  of length $N-1$ all the other positions on the lattice, ordered in function of the distance. To be more precise, the pile 
will contain the difference of positions between the point $i$ and $j$. The pile ${\cal P}_1$ will be the same for any $i$, so this operation is done only once.
Next we consider all $n(r)$ spins at a distance $r$. This is a very fast operation if we build 
a second pile ${\cal P}_2(r)$ containing the position of the first spin in ${\cal P}_1$ at a distance $r$. Then $n(r) = {\cal P}_2(r') - {\cal P}_2(r)$, with $r'$ the smallest distance on the lattice such that $r' > r$. For a given configuration of spins and 
a given starting point $i$, we need only to consider the spins with the same value 
as $S_i$ and which are not already part of the cluster under construction. We denote by $n'(r) \leq n(r)$ the number of such spins. Then the decomposition  
\bea
\label{dec}
1&=& ((1-p(r))+p(r))^{n'(r)} \\ 
 &=& p(r)^{n'(r)} + n'(r) p(r)^{n'(r)-1} (1-p(r)) \nn\\
 & & + \cdots + (1-p(r))^{n'(r)}\nn \\
&=& \sum_{k=0}^{n'(r)}  {n'(r)! \over (n'(r)-k)! k!} p(r)^{n'(r)-k} (1-p(r))^{k}  \nn
\eea
will contain the probability of having zero spins connected $p(r)^{n'(r)}$, of having one spin connected $n'(r) p(r)^{n'(r)-1} (1-p(r))$, two spins connected $n'(r) (n'(r)-1) p(r)^{n'(r)-2} (1-p(r))^2/2$, \etc\  Next we randomly generate a number $\epsilon$ in the interval $[0,1]$. If  
 $\epsilon < p(r)^{n'(r)}$, no spin will be reversed at the distance $r$. We can then ignore all the $n(r)$ spins at distance $r$, at the cost of  generating one random number. Otherwise, if 
 $ p(r)^{n'(r)} < \epsilon < p(r)^{n'(r)}  + n'(r) p(r)^{n'(r)-1} (1-p(r))$, one spin has to be reverse among 
 the possible  $n'(r)$ spins. We then need a second random number to choose one spin among the $n'(r)$. Next if $p(r)^{n'(r)}  + n'(r) p(r)^{n'(r)-1} (1-p(r)) < \epsilon $, we continue the process until
we obtain  
\beq
\label{eq4}
\epsilon <  \sum_{k=0}^{k_{max}}  {n'(r)! \over (n'(r)-k)! k! } p(r)^{n'(r)-k} (1-p(r))^{k}
\eeq
with $k_{max}$ the number of spins that we have to reverse and thus the number of random numbers that we need to generate to choose these spins among the $n'(r)$. 
 
For each distance $r$ one needs also to compute $n'(r)$. This corresponds to the delta function 
in eq.(\ref{prob}). In principle, this means that we need to perform an additional number of operations $n(r)$ to determine $n'(r)$. In fact, in most cases, this part 
can be skipped. Indeed, it is much more convenient to first compare $\epsilon $ with $p(r)^{n(r)} <p(r)^{n'(r)}  $. Thus we will compute $n'(r)$ only if $p(r)^{n(r)} < \epsilon $, which will be very rare  for $r$ large. Then in most cases one can consider all the $n(r)$ spins in the slice at distance $r$ with the cost of generating a single random number. 

A second improvement is to consider an ensemble of successive slices with $r_1 \leq r \leq r_2$.
The decomposition (\ref{dec}) is then replaced by 
\bea
1&=& \prod_{r=r_1}^{r_2}((1-p(r))+p(r))^{n'(r)} \\
&=& \left( \prod_{r=r_1}^{r_2} p(r)^{n'(r)} \right) \left(1  + \sum_{r=r_1}^{r_2} n'(r) {(1-p(r))\over  p(r)} + \cdots \right) \nn
\label{dec2}
\eea
The choices of $r_1$ and $r_2$ are guided by efficiency. Starting from a first slice $r_1$, we will 
add slices up to $r_2$ with the condition that $\prod_{r=r_1}^{r_2} p(r)^{n(r)}$ remains close 
enough to $1$ to ensure that in most cases the randomly generated number $\epsilon $ will be such that  $\epsilon <  \prod_{r=r_1}^{r_2} p(r)^{n(r)} \leq 
\prod_{r=r_1}^{r_2} p(r)^{n'(r)}$. Again, we will first compare $\epsilon $ with $ \prod_{r=r_1}^{r_2} p(r)^{n(r)}$. 
If $\epsilon $ is smaller than this product, we can then skip all the spins in the slices 
of distance $r_1 \leq r \leq r_2$ for the  cost of generating a single random number. 
Otherwise, we have to count the number of spins $n'(r)$ in each slice $r_1 \leq r \leq r_2$ which can be connected to the cluster ({\ie\ spins with the good sign and not yet part of the cluster). This corresponds to $N_{r_1,r_2}=\sum_{r_1}^{r_2} n(r)$ operations. Now, if the condition 
$  \prod_{r=r_1}^{r_2} p(r)^{n(r)} < \epsilon < \prod_{r=r_1}^{r_2} p(r)^{n'(r)}$ is satisfied, we can skip all the spins in the slices of distance $r$ with $r_1 \leq r \leq r_2$ and with a cost of generating a single random number + $N_{r_1,r_2}$ operations. 
Otherwise, we have to reverse at least one spin and we have still some additional operations.  
It can be checked that the number of additional operations in that case is smaller 
than $N_{r_1,r_2}$.   

The choice of $r_1$ and $r_2$ can now be obtained by requiring that the probability 
of reversing one spin times the number of operations is comparable to the number 
of operations for having to reverse no spins, {\ie\ when eq.(\ref{eq4}) 
is satisfied with $k_{max}=0$. As said before, for this case, the number of operations is $\mathcal{O}(1)$.
As a first approximation, the probability of reversing one spin is
 $ P_{r_1,r_2} \simeq \sum_{r=r_1}^{r_2} n(r) (1-p(r))$.  Then the optimal choice 
 $r_1$ and $r_2$ is such that $P_{r_1,r_2} \times N_{r_1,r_2} \simeq \mathcal{O}(1)$. 
 A simple extension of this argument \cite{mp} leads to the result that the total number of operations for a system with $N$ spins grows like $\mathcal{O}(N )$.
 
Our algorithm is similar in spirit with the one developed by Luijten and Bl\"ote \cite{lb} but the implementation is rather different. In their algorithm, 
the second part of the weight (\ref{prob}) is obtained by building a cumulative bond probability. In order to perform this step efficiently, they need to 
approximate the interaction between two spins by a integral. While this approximation 
does not affect the universal critical properties, other nonuniversal quantities like the critical temperatures will be different from the one that we obtained. It was argued in  \cite{fukui} 
that Luijten and Bl\"ote algorithm requires $\mathcal{O}(N \log N ))$ operations. 

Note also that while we present here results for the case of the Ising model in two dimensions,
our algorithm is valid for any Potts model and in any dimensions. 
We will present elsewhere results for the case of the Ising model with long range interactions in three dimensions \cite{mp}.

We will now report on simulation done on a triangular lattice of linear size up to $L=5120$ 
with periodic boundary conditions. To implement these boundary conditions with long range interactions, we employed the minimum image convention. 
In order to be able to obtain a good precision, we had to accumulate a lot of statisics. For each value of $\sigma, K = \beta J $ and size $L$, we 
accumulated statistics over $10^6 \times \tau$ updates, with $\tau$ the autocorrelation time. 
The time for the update of one sample of size $5120 \times 5120$ is $\simeq 0.1$ second for 
$\sigma = 0.8$ and $ \simeq 1$ second for $\sigma = 1.8$ on a recent PC.  
With the help of the Wolff cluster algorithm, the autocorrelation times are rather small even 
for the largest sizes considered. For $L=5120$, we determined $\tau \simeq 200$  for $\sigma=0.8$ and $\tau \simeq 66$  for $\sigma=1.8$. The total 
computing time required to produce the data  presented in this work corresponds to $\mathcal{O}(100)$ years of CPU time on a single core processor. In order to 
test the universality of our results and to have a more direct comparison with the results of Luijten and Bl\"ote \cite{lb}, we also performed a simulation 
for $\sigma=1.75$ on a square lattice. 

For each value of $\sigma$, we first had to determine the critical value $K_c$. This 
was done by considering a magnetic cumulant similar to the Binder cumulant \cite{Binder} and which is an adimensional quantity. 
We will consider the magnetic cumulant defined by
$B(L,K)$ :
\beq
B(L,K) = {\langle m^2\rangle^2 \over \langle m^4\rangle} \; ,
\eeq
with $\langle m^i\rangle$ the thermal average  of the magnetization to the power $i$.
This cumulant will converge to the value $1$ in the ferromagnetic phase and $1/3$ for the paramagnetic phase. The curves describing the cumulant 
versus $K$ for different sizes $L$ will cross at a value $K_c(L)$ which will converge toward the real critical point in the large size limit. 
At the crossing point, the cumulant $B(L)$ will converge towards a finite value. 
For the critical Ising model with short range interaction, the limiting 
value was computed on the triangular lattice with a value of $\lim_{L\rightarrow \infty}B(L,K_c) = 0.85872528(3)$ \cite{kb}. 
\begin{figure}
\begin{center}
\epsfxsize=255pt{\epsffile{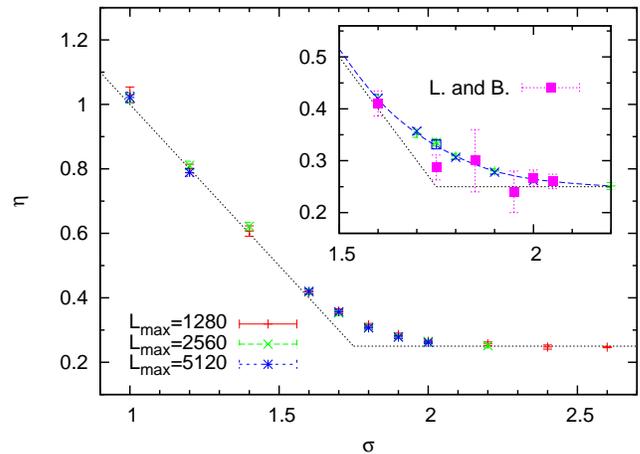}}
\end{center}
\caption{(Color online)  Exponent $\eta$ vs. $\sigma$ computed with data up to size $L=1280, 2560$ and $5120$. The inset contains a magnified part for 
$1.5 \leq \sigma \leq 2.2$ with in addition the data obtained by Luijten and 
Bl\"ote  \cite{lb}. The dotted lines correspond to the prediction from the RG analysis,  
$\eta~=~max(2-\sigma,1/4)$. The dashed line connecting our measured points 
is a guide to the eye.
}
\label{edf}
\end{figure}
Close to the critical point, the finite-size scaling behavior is expected to be of the form 
\beq
B(L,K) \simeq  f((K-K_c) L^{1/\nu}) \; ,
\eeq
with $f(x)$ a dimensionless function. 
Then the crossing of the curve $B(L,K)$ as a function of $K$ for two different sizes 
$L$ and $L'$ takes place at $K=K_c$. In the following, we will always choose $L' = 2 L$ and 
express the computed quantities as a function of $L$ only.
In practice, due to corrections to scaling, we need to take into account  additional 
correction terms to $B(L,K)$. We will consider in the following the leading correction 
of the form :
\beq
\label{eq7}
B(L,K) \simeq  f((K-K_c) L^{1/\nu}) + A_1 L^{-y_1}\; .
\eeq
With such a term, the crossing of the curves will take place at a size dependent 
$K_c(L)$ :
\beq
\label{eq8}
K_c(L) = K_c + \alpha_1 L^{-w_1} \; ,
\eeq
with $\alpha_1$ a constant and $w_1 = y_1 + 1/\nu$. 
We will then determine the value of $K_c$ by extrapolating the 
measured values of the crossing $K_c(L)$ while including one single correction. 
We have checked that with such a procedure, the values of $w_1$ 
and $K_c$ are stable if one keeps only large size data, $L \geq 100$. 
The same will be true for the computed values of $\eta$, see below. 
We have also tried to include further correction terms in (\ref{eq7}).  We have checked 
that the addition of subdominant corrections does  not affect the results 
for $L_{max} \geq 1280$, {\it i.e.} the change in $K_c$ is much 
smaller than the error bars on this value.
We also performed a fit of $B(L,K)$ close to $K_c$ with a development 
in powers of $(K-K_c)L^{1/\nu}$ and with terms $L^{-y_i}$ corresponding to 
correction to scaling. 
The obtained exponents are always in good agreement with the ones from a direct fit of the form (\ref{eq7}). 
In particular, we obtain that $\nu$ remains very close to $1$. We quote $\nu=0.96 (2)$ for $\sigma=1.6$. 
Finally, in \cite{lb}, it was claimed that logarithmic corrections to scaling were needed in order to obtain a consistent analysis  for $\sigma = 1.75$. 
This is not the case in the analysis of our data. In fact, even if we impose the existence of logarithmic corrections, we observe that most of our 
results are not affected. These corrections will mostly affect the exponents $y_i$ without a measurable change on the exponent $\eta$ for 
the largest sizes that we can simulate \cite{mp}.

In the following, we will be interested in the region $\sigma \geq 1.4$ 
since this is where we obtain new results. 
Our main results are shown in Fig.~\ref{edf}. It contains our numerical results 
for the exponent $\eta = 2 \beta/\nu$  versus $\sigma$ obtained for three ranges 
of linear sizes, $L_{max}=1280, 2560$ or $5120$. In each case, $L_{max}$ corresponds to the 
maximum sizes that we employed in order to determine the value of $K_c$ as 
described in the previous section. Then we determined the value $\eta(K_c)$ by doing 
an extrapolation from the effective exponent $\eta$ obtained from the data with $L_{max}/2$ and $L_{max}$. 
%

The values of $\eta$ that we obtained are reported in 
Table~1. We also report the value obtained for $\sigma=1.75$ on the square lattice and this value 
extrapolates well with the ones on the triangular lattice.
In this table, we can see that there is near no dependance in the size $L_{max}$. 
This confirms that our determination of $K_c$ is not affected by further corrections. 
\begin{table}
\label{Table1}
\begin{center}
\begin{tabular}{| c ||  c |  c |  c | c |}
\hline
  $\sigma$           &  $\eta(L = 1280)$    & $\eta(= 2560)$    & $\eta(L=5120) $ &  $max(2-\sigma,{1\over 4})$ \\ \hline 
 1.4   & 0.607  (11) & 0.619 (12) & -                 &   0.600      \\
 1.6   & 0.417  (6)   & 0.418 (9)   & 0.420 (8)    &   0.400      \\
 1.7   & 0.359  (7)   & 0.353 (8)   & 0.357 (7)    &   0.300             \\
 1.75 & 0.346  (7)   & 0.335  (7)  & 0.332 (8)  &   0.250              \\
 1.8   & 0.315  (5)   & 0.309 (5)   & 0.307 (5)    &   0.250              \\
 1.9   & 0.286  (5)   & 0.280 (5)   & 0.279  (5)   &   0.250              \\
 2.0   & 0.265  (3)   & 0.265 (4)   & 0.262 (4)    &   0.250              \\
 2.2   & 0.257  (6)   & 0.251 (7)   & -                 &   0.250              \\
\hline
\end{tabular}
\end{center}
\caption{ Exponent $\eta$ for $\sigma=1.4 - 2.2$. The second to fourth column contain
the values obtained by a fit with data up to linear size $L=1280, 2560$ and $5120$. The last column 
contains the predictions from the RG analysis, $\eta = max(2-\sigma, 1/4)$.}
\end{table}

From Fig.~\ref{edf}, we see that in the classical regime and in the intermediate regime up to 
$\sigma \simeq 1.5$ our results are in agreement with the prediction $\eta  = 2 - \sigma$ 
(represented by a dotted line), confirming the results of Luijten and Bl\"ote  \cite{lb}. 
For $\sigma > 2$, $\eta$ is in perfect agreement with the value 
for a short range model $\eta_{sr}  = 1/4$ (represented also by a dotted line). In the 
remaining part for $1.6 \leq \sigma \leq 2$ and shown in the inset of Fig.~\ref{edf}, 
our results do not agree with the prediction of the RG analysis \cite{fisher,sak}. 
On the contrary, we observe a set of values which interpolate smoothly between these 
two behaviors. The errors that we obtain on the measured values of $\eta$ are much 
smaller than the deviation observed to the form $max(2-\sigma,\eta_{sr})$. 
In particular, for $\sigma = 1.8$, the deviation corresponds to near $10$ times the standard deviation, {\it i.e.} $\eta - \eta_{sr} \simeq 10 d \eta $. In the inset of Fig.~\ref{edf}, 
we also show the results of Luijten and Bl\"ote from \cite{lb}. 
We note that their results are compatible with ours  but since our error bars are 
one order of magnitude smaller than theirs, we are able to observe a deviation from the RG predictions.
In fact, we can observe that Luijten and Bl\"ote results are compatible with both the RG predictions and our results \cite{deviation}. 

In this letter we obtained strong evidences that for the long range interaction 
Ising model in 2d, the exponent $\eta$ does not behave as 
predicted by RG studies in \cite{fisher,sak}, in particular in the intermediate 
regime and close to the boundary with the short range regime as shown in 
Fig.\ref{edf} and in Table~1. We believe that further studies 
are needed to recheck the RG analysis. In particular, it seems that the wave function normalization is not trivial for the long range 
interaction and this could then give further contributions which were not taken in account in the previous studies \cite{pr}.

\acknowledgments I thank M.~Rajabpour for many valuable discussions and advices and T.~Blanchard for 
a critical reading of the manuscript.


\begin{thebibliography}{99}
\bibitem{deviation} The compatibility is perfect except for $\sigma = 1.75$ for which 
Luijten and Bl\"ote result differ by $1.5 $  standard deviation compared 
to either our result or the RG prediction.
\end{thebibliography}
\end{document}